\documentstyle[nato]{crckapb}
\def\ba{\begin{eqnarray}}\def\ea{\end{eqnarray}}
\def\nn{\nonumber\\}
\begin{opening}
\title{The inelastic photon-electron  collisions with polarized beams}
\author{S.R.~Gevorkyan$^{1,2}$,E.A.Kuraev$^1$}
\institute{$^{1}$Joint Institute for Nuclear Research,141980, Dubna,Russia\\
$^{2}$ Yerevan Physics Institute, 375036, Yerevan, Armenia}
\end{opening}
\begin{document}
\runningtitle{THE CRCKAPB STYLE FILE}
\begin{abstract}
We discussed the photoproduction of pair of charged particles
$a\bar{a}\quad (a=e,\mu,\pi)$ as well as the double photon
emission processes  off an electron accounting for the
polarization of colliding particles. In the kinematics when all
the particles can be considered as a massless we obtain the
compact analytical expressions for the differential cross sections
of these processes. As the application of obtained results the
special cases of production by circular and linear polarized
photons are carry out.
\end{abstract}
We consider the lowest order inelastic QED processes in
photon-electron interaction at high energies.We take into account
the polarization of colliding photons and electron and consider
the experimental setup,when the polarization of final particles is
not measured.The interest to such kind processes is twofold.
Linear $e^+e^-$ high energy colliders(planned to be arranged~
\cite{TESLA,GKPST}) provide the possibility (using the backward
laser Compton scattering) to obtain the high energy
photon-electron colliding beams. The problem of calibration as
well as the problem of important QED background are to be taken
into account for this kind of colliders.The second important
reason to investigate such reactions is the well known possibility
to use the photoproduction of leptons pairs as a polarimeter
process(see,for instance,~\cite{Aku} and references therein) .\\
 Despite the fact that at high energy the bulk of particles
 are produced at very small angles, experimentally it is much easy to
 detect the particles produced at large angles.Thus we investigate the
kinematic in which all invariants determining  the matrix elements
of the processes under consideration will be much bigger than
masses of particles involved in the reaction.\\
We consider the following set of inelastic reaction
 \ba
 \gamma(k,\lambda_\gamma)+e(p,\lambda_e)\to
e(p',\lambda\prime_e)+a(q_-,\lambda_-)+
\bar{a}(q_+,\lambda_+);\quad a=e,\mu,\pi \nn
\gamma(k,\lambda_\gamma)+e(p,\lambda_e)\to
e(p',\lambda_e\prime)+\gamma(k_1,\lambda_1)+ \gamma(k_2,\lambda_2)
\ea
Here $\lambda_i$ are the particle helicities \\
We will work in the kinematic when all the 4-vector scalar
products defined by:
 \ba
s&=&2pp',\quad s_1=2q_-q_+,\quad t=-2pq_-,\quad t_1=-2p'q_+, \quad
u=-2pq_+, u_1=-2p'q_-,\nn \chi&=&2kp,\quad \chi'=2kp',\quad
 \chi_j=2k_jp,\quad \chi_j'=2k_jp'\quad j=1,2
\ea
 are large compared with all  masses:
 \ba
s&\sim& s_1\sim -t\sim -t_1\sim -u\sim -u_1\sim
\chi_j\sim\chi_j'>>m^2;\nn p^2&=&p^{{'}2}=q_\pm^2=k^2=k_j^2-0. \ea
To obtain the cross sections of above reactions  it is convenient
to work with helicity amplitudes of corresponding
processes~\cite{CB}. The helicity amplitudes
$M^{\lambda_-\lambda_+\lambda'}_{\lambda_\gamma\lambda}$ for
lepton pair photoproduction,
$M^{\lambda'}_{\lambda_\gamma\lambda}$ for a pair of charged pion
production and
$M^{\lambda_1\lambda_2\lambda'}_{\lambda_\gamma\lambda}$ for a two
photon final state are defined as a usual matrix elements
calculated with chiral states of photons and leptons.
\\ The square of matrix element summed over the spin states of the
final particles have the form of the conversion of the chiral
matrix with the photon density matrix \ba
 \sum |M|^2=\frac{1}{2}Tr\begin{pmatrix} {m_{11} & m_{12}
\nn  m_{21} & m_{22}}
\end{pmatrix}
\begin{pmatrix}
 {1+\xi_2 & i\xi_1-\xi_3 \nn -i\xi_1-\xi_3  & 1-\xi_2}
\end{pmatrix},
\ea with the  photon polarization vector
$\vec{\xi}=(\xi_1,\xi_2,\xi_3)$ parameterized by Stokes parameters
fulfilling the condition  $\xi_1^2+\xi_2^2+\xi_3^2\le 1$.\\
The matrix elements of the chiral matrix $m_{ij}$ are constructed
from the chiral amplitudes of the process
$M^{\lambda_-\lambda_+\lambda'}_{\lambda_\gamma\lambda_e}$ as \ba
m_{11}&=&\sum_{\lambda_-\lambda_+\lambda '}
\big|M^{\lambda_-\lambda_+\lambda '}_{++}\big|^2,\quad
m_{22}=\sum_{\lambda_-\lambda_+\lambda '}
\big|M^{\lambda_-\lambda_+\lambda '}_{-+}\big|^2, \nn
m_{12}&=&\sum_{\lambda_-\lambda_+\lambda '}
M^{\lambda_-\lambda_+\lambda'}_{++}
\big(M^{\lambda_-\lambda_+\lambda'}_{-+}\big)^*,\quad
m_{21}=m_{12}^*.
 \ea
  We put here only half of all chiral
amplitudes which correspond to $\lambda_e=\frac{1}{2}$. The other
half can
be obtained from these ones by a space parity operation. \\
The helicity amplitudes can be obtain using the relevant
representations for photon polarization vector and Dirac spinors
~\cite{BGGK}.As a result one  obtains for the elements of chiral
matrix for the processes under
consideration the following expressions: \\
1.Photoproduction of muon pair
 \ba
m_{11}&=&\frac{2w}{ss_1}(u^2+t^2),\quad
m_{22}=\frac{2w}{ss_1}(u_1^2+t_1^2), \\\nonumber
m_{12}&=&-\frac{4(w_1-iAw_2)}{(ss_1)^2}\Big[(ss_1)^2+
(tt_1)^2+(uu_1)^2- 2tt_1 uu_1-ss_1(tt_1+uu_1)+4i(tt_1-uu_1)A\Big]
\ea where \ba A&=&\epsilon_{\mu\nu\rho\sigma} q_+^{\mu} q_-^{\nu}
p^{\rho} {p'}^{\sigma},\nn
w&=&-\big(\frac{q_+}{kq_+}-\frac{q_-}{kq_-}+ \frac{p}{k
p}-\frac{p'}{k.p'}\big)^2, \nn
w_1&=&\frac{w}{2}-\frac{4s_1}{\chi_+\chi_-}+\frac{\chi\chi\prime}{4s}
\Big[\frac{w}{2}-\frac{2s}{\chi\chi\prime}-
\frac{2s_1}{\chi_+\chi_-}\Big]^2,\nn w_2&=&
\frac{2(\chi_++\chi_-)}{s\chi_+\chi_-}[\frac{w}{2}+\frac{2s}{\chi\chi\prime}-
\frac{2s_1}{\chi_+\chi_-}] \ea
 2.Electron pair photoproduction
\ba
m_{11}&=&\frac{2w}{ss_1tt_1}\Big[t^3t_1+u^3u_1+s^3s_1\Big],\quad
m_{22}=\frac{2w}{ss_1tt_1}\Big[t_1^3t+u_1^3u+s_1^3s\Big],\\\nonumber
m_{12}&=&\frac{4(w_1-iAw_2)}{(ss_1tt_1)^2}\Big[(uu_1)^2-(ss_1)^2-(tt_1)^2
\Big]\\\nonumber
&\times&\Big[\frac{1}{2}((uu_1)^2+(ss_1)^2+(tt_1)^2-2uu_1(ss_1+tt_1))+
2iA(uu_1-ss_1-tt_1)\Big] \ea

3. Photoproduction of pions \ba
 m_{11}&=&\frac{w}{2}t u,\quad
m_{22}=\frac{w}{2}t_1u_1,\\\nonumber
m_{12}&=&\frac{w_1-iAw_2}{ss_1}\Big[\frac{1}{2}(uu_1-tt_1)^2
-ss_1(uu_1+tt_1)+2i(tt_1-uu_1)A\Big] \ea The differential cross
section of any pair production has the form \ba
 \frac{d\sigma}{d\Gamma}&=&\frac{\alpha^3}{2\pi^2\chi}
\Big[m_{11}+m_{22}+\xi_2\lambda_e(m_{11}-m_{22})
-2\xi_3\mathrm{Re}(m_{12})+2\xi_1\mathrm{Im}( m_{12})\Big],\nn
d\Gamma&=&\frac{d^3p'}{\epsilon\prime}\frac{d^3q_-}{\epsilon_-}
\frac{d^3q_+}{\epsilon_+}\delta^4(p+k-p'-q_+-q_-) \ea where
$\xi_i$ and $\lambda_e$ are Stocks parameters and target electron
helicity.\\
The expressions (6)-(10) allows one to calculates the differential
cross section of the processes of pair photoproduction off an
electron with any polarization of colliding particles.  Now it is
a simple task to obtain from this expressions the particular
cases.\\Let us consider the charged particles pair production with
circular and linear polarization of photon from unpolarized
target. The differential cross section for pair production by
circularly polarized photons up to the factor $\xi_2$-the degree
of circular (left or right) photon polarization coincide with the
cross section in unpolarized case:
 \ba
\frac{d\sigma_{L(R)}}{d\Gamma}&=&
\frac{\alpha^3}{\pi^2\chi}\xi_{2L(2R)}wZ_{a\bar{a}},\nn
Z_{e\bar{e}}&=&\frac{ss_1(s^2+s_1^2)+tt_1(t^2+t_1^2)+uu_1(u^2+u_1^2)}
{ss_1tt_1}, \nn
Z_{\mu\bar{\mu}}&=&\frac{t^2+t_1^2+u^2+u_1^2}{ss_1},\quad
Z_{\pi\bar{\pi}}=\frac{tu+t_1u_1}{ss_1} \ea Two of these
quantities $Z_{e\bar{e}},Z_{\mu\bar{\mu}}$ can be obtained from
the relevant quantities obtained in papers \cite{CB} applying
the crossing symmetry transformation.\\
Much more interesting is the case of pair photoproduction by
linear polarized photons for which the Stokes parameters are
$\xi_2=0,\xi_1^2+\xi_3^2=1$.In this specific case the square of
full matrix element can be represented as the sum of diagonal term
and non diagonal one, where the polarization parameters enter only
the non diagonal term in the following form
 \ba
Re\Big((\xi_3+i\xi_1)(w_1-iAw_2)(T_1+iAT_2)\Big)=\xi_3\Big(w_1T_1+w_2T_2A^2\Big)+
\xi_1\Big(w_2T_1-w_1T_2\Big)A \ea where the
structures $T_{1(2)}$ depend on the type of created pair: \\
1. For the case of $ e^-e^+$ production \ba
T_1&=&4+8\frac{(uu_1-ss_1)(ss_1+tt_1)}{ss_1tt_1}-4(\frac{tt_1-uu_1}{ss_1})^2,
 \nn
T_2&=&8[\frac{tt_1-uu_1}{s^2s_1^2}+\frac{1}{ss_1}-\frac{2}{tt_1}]
 \ea
 2.In the case of $\mu_+\mu_-$ pair production
 \ba
T_1&=&8\frac{tt_1+uu_1-ss_1}{ss_1}-8(\frac{uu_1-tt_1}{ss_1})^2,\nn
T_2&=&16\frac{tt_1-uu_1}{(ss_1)^2}
 \ea
 3. For the case of pion pair
production: \ba
T_1&=&4(\frac{uu_1-tt_1}{ss_1})^2-4\frac{uu_1+tt_1}{ss_1},\nn
T_2&=&\frac{uu_1-tt_1}{(ss_1)^2}. \ea The same consideration can
be done for double Compton effect.The differential cross section
for this reaction in the general case, when the both primary
particles are polarized can be cast in the following form: \ba
\frac{d\sigma}{d\Gamma_\gamma}&=&\frac{\alpha^3s} {(2\pi)^2 D}
\Big[\chi\chi' ({\chi}^2+{\chi'}^{2})+
\chi_1\chi_1'(\chi_1^2+{\chi_1'}^2)+\chi_2\chi_2'(\chi_2^2+{\chi_2'}^2)
 \nn
&+&4\xi_1A(\chi_1\chi_2'-\chi_2\chi_1')-\xi_3(\chi_1\chi_2'+
\chi_2\chi_1')(\chi\chi'-\chi_1\chi_1'-\chi_2\chi_2') \nn
&+&\lambda_e\xi_2[\chi\chi'{(\chi'}^2-\chi^2)+\chi_1\chi_1'(\chi_1^2-
{\chi_1'}^2)+\chi_2\chi_2'(\chi_2^2-{\chi_2'}^2)]\Big] \ea with
$D=\chi^2\chi'\chi_1\chi_1'\chi_2\chi_2'$ and
$A=\epsilon_{\mu\nu\rho\sigma}k_2^\mu k_1^\nu p^\rho
{p\prime}^\sigma$.\\
For unpolarized target  the cross section of double Compton effect
in the case of circularly polarized photon is:
 \ba
d\sigma_{L(R)}&=&\xi_{L(R)}\frac{2s\alpha^3}{\pi^2D} Z_\gamma
d\Gamma_\gamma,\nn Z_\gamma&=& \chi\chi'(\chi^2+\chi^{{'}2})+
\chi_1\chi_1'(\chi_1^2+\chi_1^{{'}2})+\chi_2\chi_2'(\chi_2^2+\chi_2^{{'}2})
 \ea with \ba
d\Gamma_\gamma=\frac{d^3p_1'}{2\epsilon_1'}\frac{d^3k_1}{2\omega_1}
\frac{d^3k_2}{2\omega_2}\delta^4(k+p_1-p_1'-k_1-k_2). \ea The
summed on spin states square of the matrix element of the double
Compton scattering process can as well be obtained from the ones
for three photon annihilation of electron-positron pair,derived in
~ \cite{CB}. \\
For the case of double Compton effect under the linearly polarized
photons one can obtain for non diagonal part of cross section
which as in above case depend on Stoke's parameters the following
expressions:  \ba &Re&\Big((\xi_3+i\xi_1)(T_1+iAT_2)\Big)=\xi_3T_1
-\xi_1AT_2\nn
&T_1&=\frac{16s}{\chi\chi_1\chi_2\chi'\chi_1'\chi_2'}
(\chi_1\chi_2'+\chi_2\chi_1')[\chi_1\chi_2'+\chi_2\chi_1'+s(s-\chi+\chi')],
\nn &T_2&=\frac{8s}{\chi\chi_1\chi_2\chi'\chi_1'\chi_2'}
(\chi_1\chi_2'-\chi_2\chi_1') \ea

\end{document}